\documentclass[%
preprint, 
superscriptaddress,
groupedaddress,
unsortedaddress,
runinaddress,
frontmatterverbose, 
preprintnumbers,nofootinbib,
nobibnotes,
bibnotes,
amsmath,amssymb,
aps, physrev,
prstab,
prstper,
floatfix,
]{revtex4}

\usepackage{graphicx}  
\usepackage{dcolumn}   
\usepackage{amsmath}   
\usepackage{amssymb}   
\usepackage{comment}
\usepackage[subrefformat=parens,labelformat=parens,caption=false]{subfig}
\usepackage[latin1,utf8]{inputenc}
\usepackage{color}
\usepackage{slashed}
\usepackage{footnote}
\usepackage[colorlinks=true,
            urlcolor=blue,
            anchorcolor=blue,
            citecolor=blue,
            filecolor=blue,
            linkcolor=blue,
            menucolor=blue,
            linktocpage=true,
            pdfproducer=medialab,
            pdfa=true,
            bookmarks=false]{hyperref}
\usepackage{paralist}
\usepackage{url}
\usepackage{cleveref}

\makeatletter
\def\@makefnmark{\hbox{\textsuperscript{\normalfont\@thefnmark}}}
\makeatother

\begin{document}
% 
%\preprint{WITS-MITP 019}
% 
\title{Role of angular observables in probing non-standard $HZZ$ couplings at an electron-proton collider}
\author{Pramod Sharma}
\email{pramodsharma.iiser@gmail.com}
\affiliation{Indian Institute of Science Education and Research, Knowledge City, Sector 81, S. A. S. Nagar, Manauli PO 140306, Punjab, India.}\author{Ambresh Shivaji}
\email{ashivaji@iisermohali.ac.in}
\affiliation{Indian Institute of Science Education and Research, Knowledge City, Sector 81, S. A. S. Nagar, Manauli PO 140306, Punjab, India.}
% 
% \date{\today}

\begin{abstract}
	In this study, we explore various lab-frame angular observables at Large Hadron-electron Collider (LHeC) to test their sensitivity on the most general structure of the Higgs ($H$) to neutral weak boson ($Z$) coupling ($HZZ$) via the process $e^- p \rightarrow e^- H j$ at the center of mass energy $\sqrt{s}~\approx$ 1.3 TeV. The most general Lorentz structure of the $HZZ$ coupling beyond the standard model is composed of two CP-even ($\lambda_{1Z}$ and $\lambda_{2Z}$) and one CP-odd (${\tilde \lambda_Z}$) components. In a previous study, we looked at the absolute value of azimuthal correlation ($|\Delta \phi|$) between the final state electron and the jet to derive constraints on the non-standard $HZZ$ couplings. This choice of observable is motivated by its potential to discriminate between CP-odd and CP-even couplings in the $e^- p \rightarrow \nu_e H j$ process. Since the process $e^- p \rightarrow e^- H j$ has an electron in the final state, one can construct more angular observables. In addition to $|\Delta \phi|$, we identify new angular observables: the sign-sensitive azimuthal correlation ($\Delta \phi$) and the polar angle of the final state electron ($\theta$), suitable for constraining non-standard $HZZ$ coupling.
    Our analysis shows that these new angles improve the constraints on $\lambda_{2Z}$ and ${\tilde \lambda}_Z$ in the range of 48-67\%. We also study the potential of the asymmetry in $\Delta \phi$ to constrain the parameters corresponding to the CP-odd coupling.
\end{abstract}

\maketitle

%%%%%%%%%%%%%%%%%%%%%%%%%%%%%%%%%%%%%%%%%%%%%%%%%%%%%%%%%%%%%%%%%%%%%%%%%%%%%%%
% 
\section{Introduction}
The discovery of a fundamental scalar of mass 125 GeV  at the Large Hadron Collider(LHC) 
is consistent with the Higgs boson ($H$) predicted in the minimal electroweak standard model. 
Its characterization by the ATLAS and CMS collaborations \cite{ATLAS:2012yve,CMS:2012qbp} has significantly improved 
our understanding of its coupling with other standard model particles. 
In the kappa framework, the coupling of $H$ with the $Z$ boson has been measured 
with an accuracy below $7\%$ using LHC Run 2 data in CMS \cite{CMS:2022dwd} and ATLAS~\cite{ATLAS:2022vkf} experiments, and it is expected to reach 1.5\% at HL-LHC~\cite{Cepeda:2019klc} and 1.2\% at LHeC~\cite{LHeC:2020van}. The combined results from LHeC+HL-LHC will bring it down to $\sim 0.82 \%$~\cite{LHeC:2020van}. There is a scope for potential new physics beyond the Standard Model to hide in the LHC data on the Higgs boson. Understanding the possible nature of new physics due to hidden symmetries requires going beyond the kappa framework in collider searches. In this regard, new physics parameterization based on higher dimensional operators, constructed out of standard model fields, provides a consistent theoretical framework. Such higher dimensional operators often lead to a non-trivial modification in standard model couplings. This framework is particularly suitable for capturing new physics effects that may appear at high energy scales and are hard to detect directly.

The most general $HZZ$ interaction vertex takes the form~\cite{Hagiwara1993ProbingTS,PhysRevD.78.115016,Biswal:2012mp,Alloul:2013naa,Kumar:2015kca,Hernandez-Juarez:2023dor}~\footnote{The non-standard $HVV$ terms in Eq. 1.2 of~\cite{Sharma:2022epc} should have an overall -ve sign.},
\begin{align}
	\Gamma_{HZZ}^{\mu\nu}(q_1,q_2) = & g_Z m_Z \kappa_Z g^{\mu\nu} - \frac{g}{m_W} \Big[\lambda_{1Z} (q_1^\nu q_2^\mu - g^{\mu \nu} q_1.q_2 ) \notag \\
	& + \lambda_{2Z} (q_1^{\mu}q_1^{\nu}+q_2^{\mu}q_2^{\nu}-g^{\mu \nu} q_1.q_1-g^{\mu \nu} q_2.q_2) \notag \\
	& +  \widetilde{\lambda}_Z ~ \epsilon^{\mu \nu \alpha \beta} q_{1 \alpha} q_{2 \beta} \Big].
 \label{Eq:BSMcoup}
\end{align}
In the above equation, we use the prescription $\epsilon^{0123} = 1$. Here, $(q_1, q_2)$ and $(\mu, \nu)$ represent the momenta and Lorentz indices of the two $Z$ bosons, respectively. Clearly, the standard model can be recovered from the above by setting $\kappa_Z=1$ and $\lambda_{1Z}=\lambda_{2Z}={\tilde \lambda}_Z =0$.
The presence of momentum-dependent couplings is typical of higher-dimensional operators, and high-energy particle colliders are suitable experiments to probe such momentum-dependent couplings. The collider phenomenology of couplings of Higgs boson with vector bosons given in Eq.~\ref{Eq:BSMcoup}, has been performed at LHC in ATLAS~\cite{ATLAS:2020evk,ATLAS:2023mqy} and CMS~\cite{CMS:2017len,CMS:2019jdw,CMS:2019ekd,CMS:2021nnc,CMS:2022uox,CMS:2024bua} experiments via the Higgs production in VBF, VH processes and four lepton decay channel of Higgs.  Latest constraints from ATLAS~\cite{ATLAS:2023mqy} is [-0.33, 0.48] on $\widetilde{\lambda}_Z$ at 95\% C.L. and from CMS~\cite{CMS:2024bua} on $\lambda_{1Z}$, $\lambda_{2Z}$, and $\widetilde{\lambda}_Z$ are [-0.13, 0.09], [-0.01, 0.01], and [-0.13, 0.07], respectively at 68\% C.L..

An electron-proton ($e^- p$) collider has advantage over both $e^-e^+$ and $pp$ colliders. It provides an 
opportunity to study high-energy interactions with relatively low backgrounds.
In this work, we present a cut-based analysis to probe the structure of non-standard $HZZ$ couplings given in Eq.~\ref{Eq:BSMcoup} at a proposed electron-proton 
collider~\cite{LHeCStudyGroup:2012zhm} in  $e^- + p \to e^- H j + X $ scattering process. At the parton level, this process proceeds via the so-called neutral current (NC) 
interaction in which $Z$ bosons from the incoming electron and quark fuse to produce a Higgs boson. 

One of the challenges in new physics searches at colliders is to identify observables that can efficiently distinguish among various Lorentz structures present in a coupling. In Ref ~\cite{Biswal:2012mp}, it was found that CP-even and CP-odd non-standard $HWW$ coupling structures can be probed efficiently 
in $e^- + p \to \nu_e H j + X$ (led by a charge current interaction) process using 
$|\Delta \phi_{{\slashed{E}}j}|$ angular observable defined as the azimuthal angle 
between missing transverse energy and the light jet. 
The same angular observable involving $e^-$ and $j$ has been studied in detail for the neutral current 
process as well using the BSM framework given in Eq.~\ref{Eq:BSMcoup}~\cite{Sharma:2022epc}. Unlike the case of the charged current process, in the neutral current process, we have access to the four momenta of all final-state particles.  
This situation allows us to construct new angular observables and investigate their efficiency 
in probing non-standard $HZZ$ coupling structures. The aim is to identify BSM parameter-specific angles that 
can provide the best constraints on individual parameters of Eq.~\ref{Eq:BSMcoup}.

\section{General consideration}
In this work, we do analysis at the parton level where we use \texttt{MadGraph5\_aMC@NLO} (MG5)~\cite{Alwall:2014hca} to generate signal and background events.
 The NC process was analyzed in Ref.~\cite{Sharma:2022epc} using the $|\Delta \phi_{ej}|$ distribution to constrain $HZZ$ coupling. The simulation part from that study is summarised in this section which we use to produce results in section~\ref{section:ang_obs}. We suggest the reader follow the Ref.~\cite{Sharma:2022epc} for a detailed discussion on the simulation methodology. We consider the \texttt{NN23LO1} parton distribution function with default MadGraph dynamical scales for renormalization and factorization. 
Signal events are identified with $e^-$, two $b$ jets, and one light jet in the final state of charge current process. Contributions to the background come from the processes $e^- p \rightarrow e^- b \bar{b} j$, $e^- jjj$, and $e^- b \bar{b} j j$, where $j$ represents a jet from light quark ($u, d, s, c$) and gluons. Our kinematic selections to improve the significance of the SM signal over all the backgrounds are given as follows:\\
(1) $p_T(e^-)> 20$ GeV, $p_T(b)>30$ GeV, and $p_T~ >$ 30 GeV cut on the less forward jet of 2 $jet$ background,\\
(2)  $|M_{b \bar{b}}-M_H| \leq 15$ GeV which is a window of 30 GeV around Higgs mass, \\
(3) $ |\eta(e^-)|<2.5$, $2<\eta(j)<5$, $0.5<\eta(b)<3$, \\
(4) Invariant mass of Higgs and $jet$ ($M_{Hj}$) $>$ 300 GeV. \\

Cut optimization increases the signal-to-background ratio (S/B) to 0.41. In the event selections, the tagging efficiency of a $b$-jet is 0.6. The probability of miss-tagging $c$-jet and light jet ($u,d,s,g$) as a $b$-jet are 0.1 and 0.01, respectively. Tagging and miss-tagging rates substantially reduce backgrounds with light jets in the final state. So we do not consider the 3 $jet$ background in our analysis. We also use smearing of partonic energy by formula $\frac{\sigma_E}{E} = a/\sqrt{E} \oplus b$ to include  hadronization effects at detector where 
$a=0.6,~b=0.04$ for parton jets, and $a=0.12,~b=0.02$ for electron are chosen from Ref.~\cite{LHeC:2020van}. 

\section{Angular observables}
\label{section:ang_obs}
CP even and CP odd observables for lepton and hadron colliders for couplings between standard model particles have been studied in~\cite{Han:2000mi,Hagiwara:2000tk,Biswal:2008tg,Han:2009ra,Christensen:2010pf,Desai:2011yj,Mileo:2016mxg,Belyaev:2016hak,Ferreira:2016jea,Li:2019evl,Tiwari:2019kly,Rao:2019hsp,Faroughy:2019ird,Barman:2021yfh,Rao:2022olq,Azevedo:2022jnd,Li:2024mxw}. \emph{Genuine} CP-even and CP-odd observables cannot be constructed for $ep$ collider studies since the initial state in this collider is not the CP eigenstate~\cite{Han:2009ra}. In this case, we can construct observables sensitive to the Lorentz structure of couplings. Such observables have been used in~\cite{Plehn:2001nj, Hankele:2006ma, Biswal:2012mp}. We are interested in angular observables over radial observables, as the former is known for being sensitive to the spin and parity of particles.
In this section, we discuss angular observables sensitive to the momentum-dependent structure of $HZZ$ coupling.  These distributions are produced using generation level cuts, $ p_T(j) > 10~ \text{GeV}, p_T(e) > 10~ \text{GeV}, \Delta R (e, j) > 0.4,
	 \Delta R (b, \bar{b}) > 0.4, \Delta R (e, b) > 0.4, \Delta R (b, j) > 0.4 $ in MG5.

Angular distributions defined in Eqs.~\ref{obs_theta}-\ref{obs_beta} are $\theta_e$, the angle made by scattered $e^-$ with the direction of incident $e^-$ beam, $\alpha_{ej}$, the angle between scattered $e^-$ and the $jet$, and $\beta_{ejH}$, the angle between Higgs and normal to the plane of $e^-$ and $jet$. We also consider $\Delta \phi_{ej}$ in the range $-\pi$ to $\pi$ instead of $|\Delta \phi_{ej}|$ which is defined between 0 to $\pi$ and has been studied in Ref.~\cite{Sharma:2022epc}.
We have chosen final state particles $e^-$ and $jet$ in distributions since these particles are directly related to the momenta of the legs involved in the non-standard $HZZ$ coupling. Therefore, distributions with combinations of these two particles are likely to capture the most sensitivity.
\begin{eqnarray}
\theta_{e} = \text{cos}^{-1}\left( \frac{\vec{p}_e . \hat{z}}{|\vec{p}_e|} \right) \label{obs_theta} \\
\alpha_{ej} = \text{cos}^{-1}\left( \frac{\vec{p}_e . \vec{p}_j}{|\vec{p}_e||\vec{p}_j|} \right) 
\label{obs_alpha} \\ 
\Delta \phi_{ej} = \text{tan2}^{-1}\left( \frac{p_{ey}}{p_{ex}} \right) - \text{tan2}^{-1}\left( \frac{p_{jy}}{p_{jx}} \right) 
=\text{tan2}^{-1}\left[ \frac{(\vec{p}_e \times \vec{p}_j).\hat{z}}{\vec{p}_{Te}.\vec{p}_{Tj}} \right]
	\label{obs_Dphi} \\
    \beta_{ejH} = \text{cos}^{-1}\left( \frac{(\vec{p}_e \times \vec{p}_j).\vec{p}_H}{|\vec{p}_e \times \vec{p}_j| |\vec{p}_H|} \right) = \text{cos}^{-1}\left( \frac{(\vec{p}_e \times \vec{p}_j).\hat{z}(E_1-E_2)}{|\vec{p}_e \times \vec{p}_j| |\vec{p}_H|}
    \right)
     	\label{obs_beta}
\end{eqnarray}
Here $\vec{p}_i~ (i = e, j, H)$ corresponds to the momentum of final state particles (electron, $jet$, Higgs) of the NC process. $E_1$ and $E_2$ are energies the initial states $e^-$ and quark, respectively.

\begin{figure}[h!]
\begin{center}
  	\subfloat[]{\includegraphics[width=0.45\textwidth]{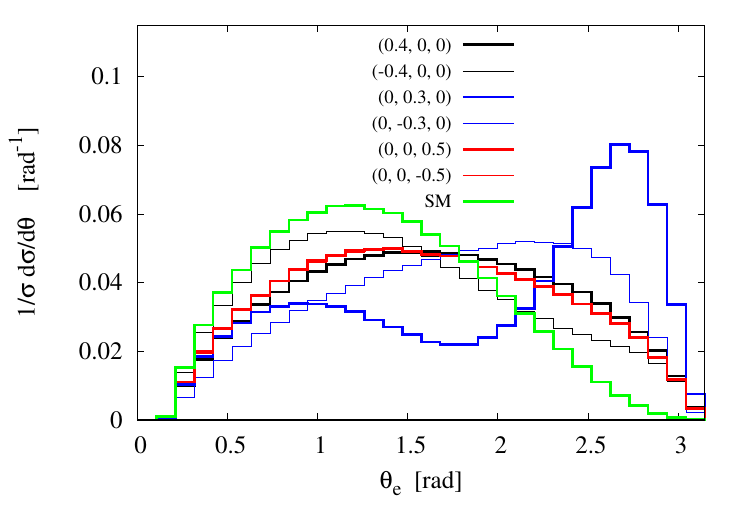}
   \label{fig_obs:theta}}
	\subfloat[]{\includegraphics[width=0.45\textwidth]{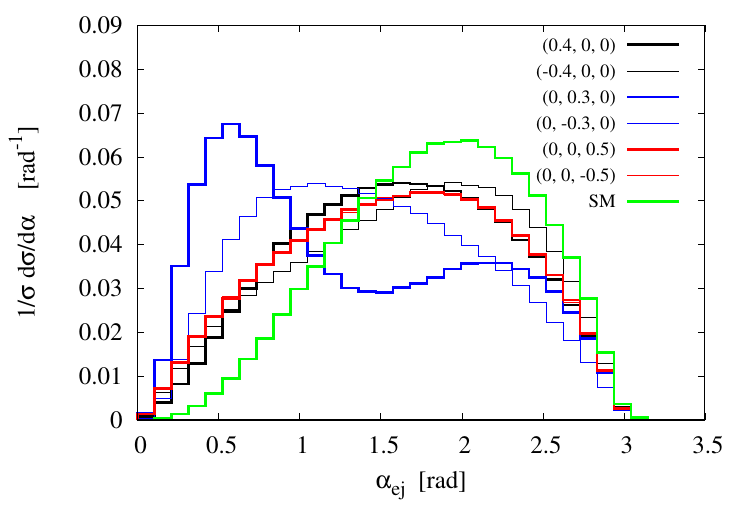}
 \label{fig_obs:alpha}}\\
	\subfloat[]{\includegraphics[width=0.45\textwidth]{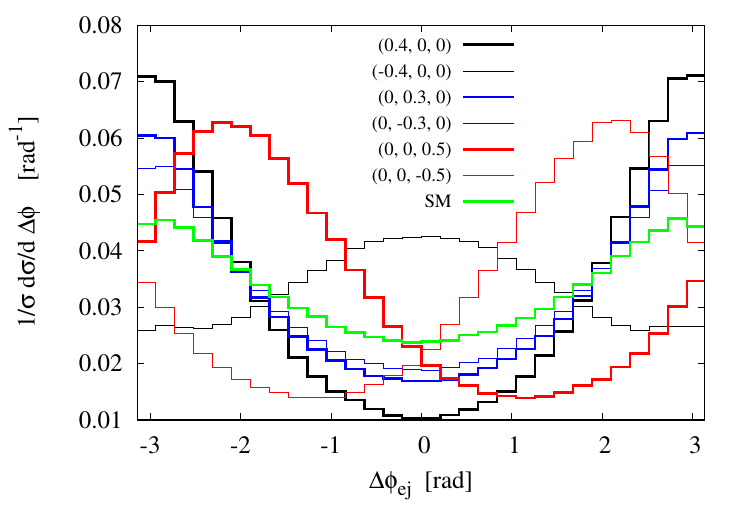}
 \label{fig_obs:Dphi}}
 	\subfloat[]{\includegraphics[width=0.45\textwidth]{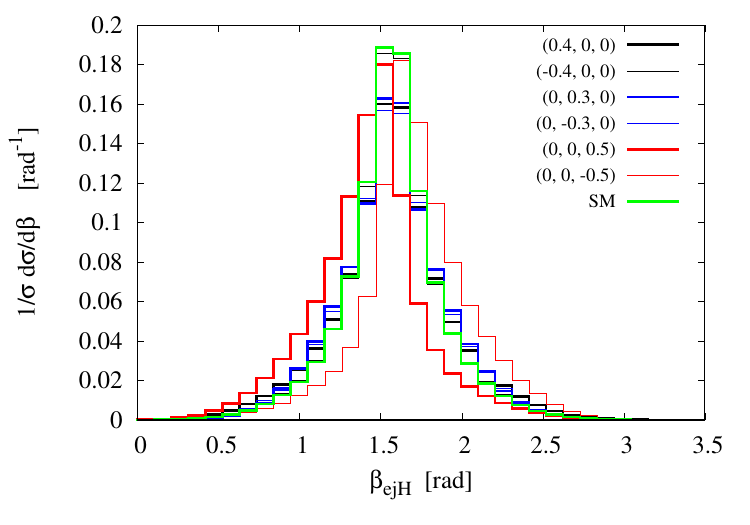}
  \label{fig_obs:beta}}
	\caption{Angular observables sensitive to BSM coupling associated with parameters ($\lambda_{1Z}, \lambda_{2Z}, \widetilde{\lambda}_Z$). }
	\label{fig_obs}
\end{center}
\end{figure}

Since the $\kappa_Z$ parameter only scales the SM coupling, it cannot capture deviations with respect to the SM in differential distributions. Therefore, we do not consider this parameter in our study.
In Fig.~\ref{fig_obs}, we show the sensitivity of the angular observables to the Lorentz structures associated with the parameters $\lambda_{1Z}$, $\lambda_{2Z}$, and $\widetilde{\lambda}_Z$. In order to test the effects of each BSM coupling on the kinematic observables, only one parameter is taken non-zero at a time. We have selected benchmark values ($\lambda_{1Z}$ = $\pm$0.4, $\lambda_{2Z}$ = $\pm$0.3, $\widetilde{\lambda}_Z$ = $\pm$ 0.5) for these parameters. These values are motivated by constraints on the BSM parameters that we obtained from a previous analysis \cite{Sharma:2022epc}. These values are only used to demonstrate deviations from the SM, while the rest of the analysis is independent of these specific choices of values of parameters. When the value of these parameters is set to zero, it replicates the SM distribution, as indicated by the green line in Fig.~\ref{fig_obs}.

\begin{figure}
    \centering
    \includegraphics[width=0.5\linewidth]{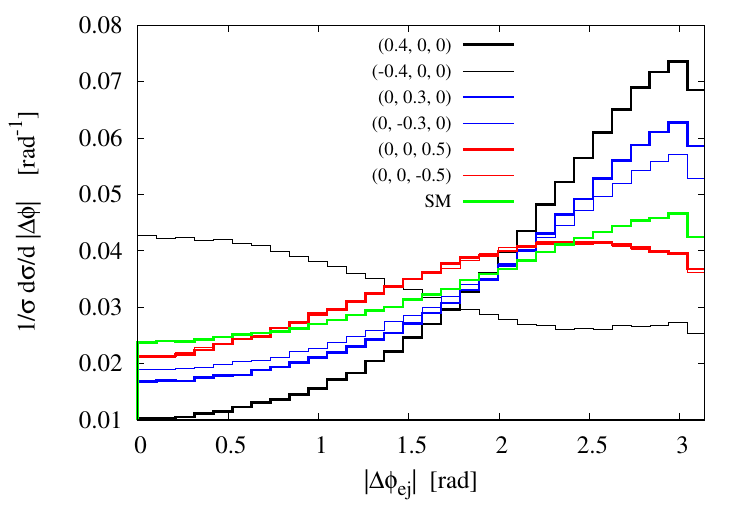}
    \caption{$|\Delta \phi_{ej}|$ distribution for benchmark values of ($\lambda_{1Z}, \lambda_{2Z}, \widetilde{\lambda}_Z$).}
    \label{fig:enter-label}
\end{figure}

As shown in Fig.~\ref{fig_obs:theta}, the distribution $\theta_e$ is more sensitive to the $\lambda_{2Z}$ parameter than to other parameters of BSM. In the case of SM, like the $jet$, the electron should be emitted along the beam directions due to the kinematics dictated by the weak boson propagator associated with the electron. However, the boost of the center-of-mass frame along the proton beam forces the electron to scatter away from its incident beam direction. Due to this, a significant number of events with $\theta_e$ away from zero also occur~\footnote{With a higher beam energy of the electron, $\theta_e$ starts peaking in the forward direction.}.  This behavior remains the same in the presence of $\lambda_{1Z}$ and $\widetilde{\lambda}_Z$. In the case of $\lambda_{2Z}$, the $\theta_e$ distribution is shifted more towards $\pi$.

Similarly, in Fig.~\ref{fig_obs:alpha}, the $\alpha_{ej}$ distribution corresponding to $\lambda_{2Z}$ is characteristically
very different from the SM prediction, while in the case of $\lambda_{1Z}$ and $\widetilde{\lambda}_Z$, it follows behavior similar to the SM contribution. This can be understood from the polar angle distributions of the electron (Fig.~\ref{fig_obs:theta}) and the jet. The jet is produced in the direction of the incident proton beam for the SM and all three BSM coupling scenarios. However, the electron is mostly transverse for $\lambda_{1Z}$, $\widetilde{\lambda}_Z$, and the SM. For these two BSM couplings, electrons in the transverse direction and $jets$ along the proton beam lead events to largely favor the angle $\alpha_{ej}$ around $\pi/2$. In the case of $\lambda_{2Z}=0.3$, the polar angle distribution of electron peaks around $\theta_e = \pi/3$, and $4\pi/5$ while for $jet$ it peaks along the proton beam direction.
%around $\theta_j = \pi$. 
The correlation between $\theta_e$ and $\alpha_{ej}$ suggests that the two observables should be equally sensitive to the BSM parameter $\lambda_{2Z}$.

The $\Delta \phi_{ej}$ distribution in Fig.~\ref{fig_obs:Dphi}, ranging from -$\pi$ to $\pi$, contains more differential information than the $|\Delta \phi_{ej}|$ distribution used in~\cite{Biswal:2012mp,Sharma:2022epc}. This motivates us to examine the $\Delta \phi_{ej}$ distribution, which can be constructed without any ambiguity~\footnote{In the case of VBF, where a similar angle between two jets can be constructed, only $|\Delta \phi_{jj}|$ is meaningful.}. The $\Delta \phi_{ej}$ distribution is defined using the $\text{tan2}^{-1}$ (quadrant inverse tangent) function, as it covers all four quadrants from $0$ to $2\pi$ of the azimuthal angle. This function properly accounts for all possible directions of outgoing particles in the transverse plane of the detector. For most of the couplings, the $\Delta \phi_{ej}$ distribution peaks at values around $\pm \pi$. The transverse momentum distribution of Higgs suggests that most of the events lie at low $p_T$ of Higgs; therefore, the electron and $jet$ pairs are likely to be produced back-to-back. However, the $\Delta \phi_{ej}$ distribution for $\lambda_{1Z} = - 0.4$ peaks at  $\Delta \phi_{ej} = 0$. 
Such behavior of the $\Delta \phi_{ej}$ distribution arises because the interference term is large and negative in the central bins of the distribution for $\lambda_{1Z}$, which results in a positive contribution in those bins for negative values of $\lambda_{1Z}$. In contrast, for $\lambda_{2Z}$, the interference term is negative and small with respect to the quadratic term across all bins. Consequently, the quadratic term primarily dictates the shape of the  $\Delta \phi_{ej}$ distribution for $\lambda_{2Z}$, as illustrated in Fig.~\ref{fig_obs:Dphi}.

The interference term for the CP-odd contribution plays a crucial role in shaping the behavior of the $\Delta \phi_{ej}$ distribution. 
The interference between the CP-odd term and the SM amplitude is given by:
\begin{eqnarray}
        \mathcal{R}e (\mathcal{M}_{
\rm SM}^{\ast} \mathcal{M}_{\widetilde{\lambda}}) &\propto& ~ \widetilde{\lambda}_Z ~ \epsilon^{\mu \nu \alpha \beta}  p_{1 \mu} p_{2 \nu} p_{e \alpha} p_{j \beta} \notag \\
%& = - \widetilde{\lambda}_Z (E_1 p_{2z} + E_2 p_{1z}) (\vec{p}_e \times \vec{p}_j).\hat{z} \\
 &\propto&   - \widetilde{\lambda}_Z ~\text{sin}\Delta\phi_{ej}
 \label{amp}
\end{eqnarray}

We can also observe in Fig.~\ref{fig_obs:Dphi} that the $\Delta \phi_{ej}$ distribution is asymmetric about $\Delta \phi_{ej} = 0$ when the CP-odd coupling $\widetilde{\lambda}_Z$ is non-zero. The interference term in Eq.~\ref{amp} does not contribute at $\Delta \phi_{ej} = 0$. We note that for +ve values of  $\widetilde{\lambda}_Z$, the interference term is a positive contribution for $\Delta \phi_{ej} < 0$ and a negative contribution for $\Delta \phi_{ej} > 0$. This leads to an asymmetry around $\Delta \phi_{ej} = 0$. The behavior of this distribution is exactly the opposite for -ve values of $\widetilde{\lambda}_Z$. 
The asymmetry also remains when one of the CP-even couplings is present along with the CP-odd coupling. This feature provides the advantage of constraining the CP-odd coupling even in the presence of CP-even couplings.

Following the discussion on vector boson fusion process at the LHC~\cite{Hankele:2006ma}, we note that the $|\Delta \phi_{ej}|$ distribution for pure CP-even and CP-odd couplings (where ``pure'' means the Standard Model contribution is not considered) exhibits distinct behaviors: it peaks at $|\Delta \phi_{ej}| = 0, \pi$ for CP-even couplings, while for CP-odd couplings, it dips at $|\Delta \phi_{ej}| = \pi/2$. In contrast, the pure SM contribution results in a relatively flat $|\Delta \phi_{ej}|$ distribution. This demonstrates that the distribution is highly sensitive to pure CP-even and CP-odd coupling structures. However, when the CP-odd coupling is present in addition to the SM contribution, the distribution becomes insensitive to the sign of the CP-odd parameter $\widetilde{\lambda}_Z$ as shown in Fig.~\ref{fig:enter-label}.
This effect arises due to the sign of the azimuthal correlation between the outgoing electron and the jet present in the interference term.

The interference term linear in the CP-odd coupling is an odd function of $\Delta \phi_{ej}$, as shown in Eq.~\ref{amp}. In contrast, for the CP-even coupling, including the SM case, the amplitudes are even functions of $\Delta \phi_{ej}$, as discussed in Refs.\cite{Biswal:2008tg,Plehn:2001nj,Hankele:2006ma}. This implies that interference with CP-odd coupling is the only term in amplitude-squared that is sensitive to the sign of  $\Delta \phi_{ej}$. When obtaining $|\Delta \phi_{ej}|$ from $\Delta \phi_{ej}$ via phase space integration, the 
interference term cancels. Therefore, the distributions presented in Ref.~\cite{Sharma:2022epc} and Fig.~\ref{fig:enter-label} in this work are insensitive to the sign of the CP-odd parameter.  However, the interference terms proportional to the CP-even parameters $\lambda_{1Z}$ and $\lambda_{2Z}$ remain intact in phase space integration and show presence in both $|\Delta \phi_{ej}|$ and $\Delta \phi_{ej}$ distributions.

Constraints on the CP-odd coupling of $HWW$ have been obtained via charged current process in Refs.~\cite{Biswal:2012mp,Sharma:2022epc} using $|\Delta \phi_{\slashed{E}j}|$. Similar to $\Delta \phi_{ej}$ given in Eq.~\ref{obs_Dphi}, one can also define the correlation $\Delta \phi_{\slashed{E}j}$, between missing energy and jet of the \texttt{CC} process, as missing energy information in the transverse plane can be reconstructed from the final state particles.

A similar argument applies to understanding the asymmetric nature of $\beta_{ejH}$ distribution.
 The $\beta_{ejH}$ distribution is related to the angle made by the normal to the $e^-$ and $ jet$ plane and the electron beam direction, as defined in Eq.~\ref{obs_beta}. Since the $jet$ is predominantly produced along the proton beam direction, the plane formed by the scattered $e^-$ and the $jet$ passes through the beam axis. Consequently, a large number of events in the $\beta_{ejH}$ distribution accumulate around $\beta_{ejH}=\pi/2$. Moreover, the $\beta_{ejH}$ and $\Delta \phi_{ej}$ distributions exhibit similarities due to their dependence on the common factor $(\vec{p}_e \times \vec{p}_j).\hat{z}$, which appears in both observables and also contributes to the amplitude for the CP-odd coupling.  As a result,  both observables provide similar sensitivity to the $\widetilde{\lambda}_Z$ coupling, making it reasonable to expect similar constraints on $\widetilde{\lambda}_Z$ from these two distributions. Due to the similarity between  $\theta_e$ and $\alpha_{ej}$,  as well as between $\Delta \phi_{ej}$ and $\beta_{ejH}$, we focus our discussion on the comparatively simpler distributions of $\theta_e$, $|\Delta \phi_{ej}|$, and $\Delta \phi_{ej}$ in the rest of the analysis.

\section{Bounds on BSM  parameters} 
We use $\chi^2$ analysis to estimate constraints on BSM parameters $\lambda_i$ assuming the standard model hypothesis.
The $\chi^2$ as a function of BSM couplings $\lambda_i$ and $\lambda_j$ (where $\lambda_i$ and $\lambda_j$ are $\lambda_{1Z}$, $\lambda_{2Z}$, and $\widetilde{\lambda}_Z$) for any observable ($\mathcal{O}$) defined in Eqs.~\ref{obs_theta}-\ref{obs_beta} is given by,
\begin{align}
	\chi^2(\lambda_{i}, \lambda_j)= \sum_{k=1}^{n} \left( \frac{\mathcal{O}_k^{\rm BSM}(\lambda_{i}, \lambda_j)-\mathcal{O}_k^{\rm SM}}{\Delta \mathcal{O}_k} \right)^2
 \label{chiSq}
\end{align}
Here, $n$ is the total number of bins in which an observable is distributed. $\mathcal{O}_k^{\rm SM}$ and $\mathcal{O}_k^{\rm BSM}$ are the values of the observables of Eqs.~\ref{obs_theta}-\ref{obs_beta} in the $k^{\rm th}$ bin for SM and BSM, respectively. $\Delta \mathcal{O}_k$ is the statistical uncertainty in the $k^{\rm th}$ bin.

\begin{figure}
	\subfloat[]{\includegraphics[width=0.40\textwidth]{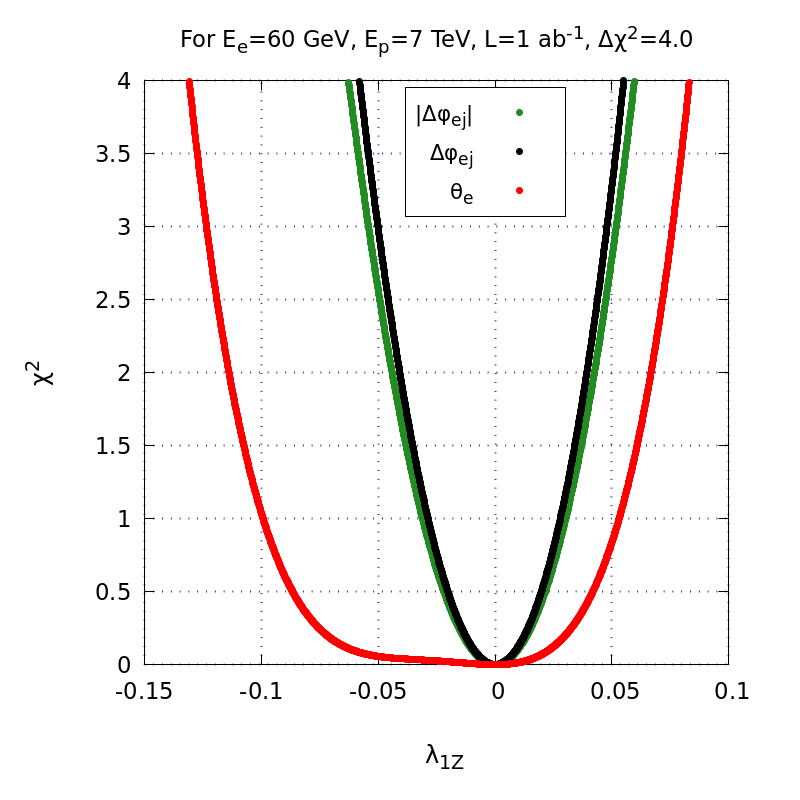}
    \label{chisq_l1}
    }
 	\subfloat[]{\includegraphics[width=0.40\textwidth]{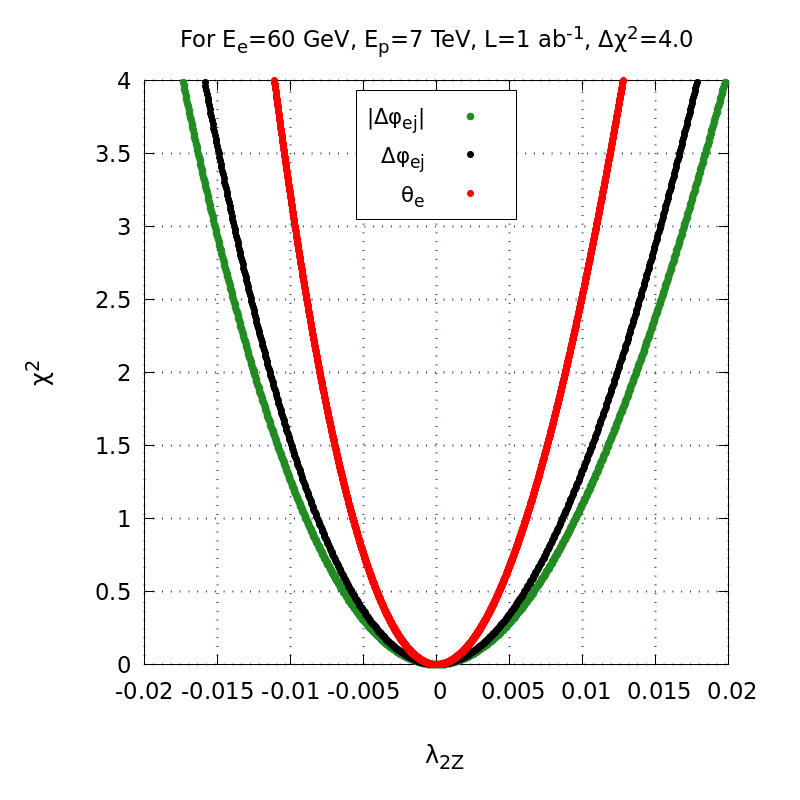}
    \label{chisq_l2}
    }\\
	\subfloat[]{\includegraphics[width=0.40\textwidth]{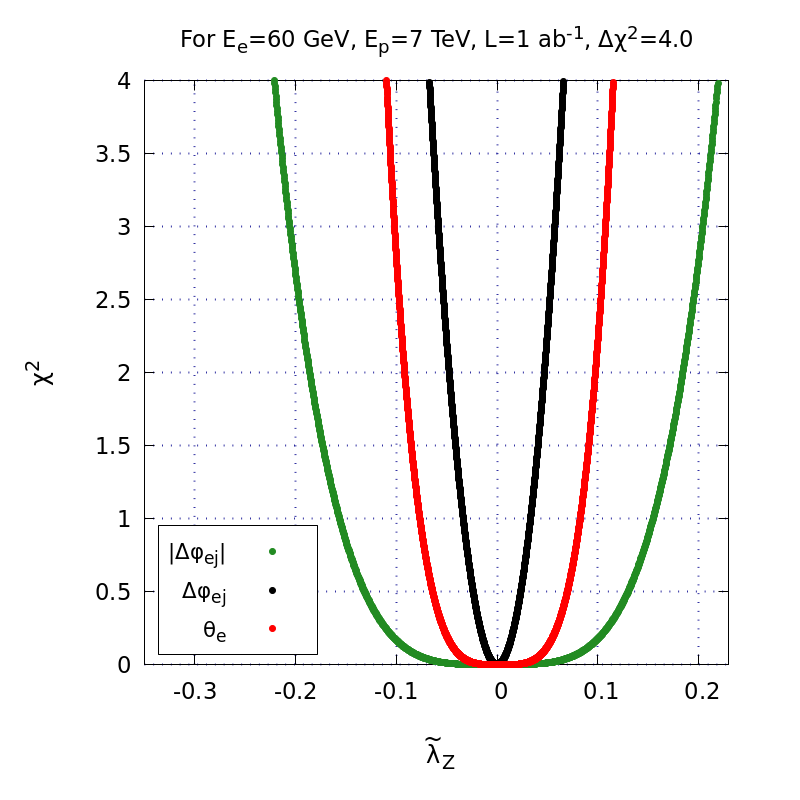}
    \label{chisq_lt}
    }
    \subfloat[]{\includegraphics[width=0.40\textwidth]{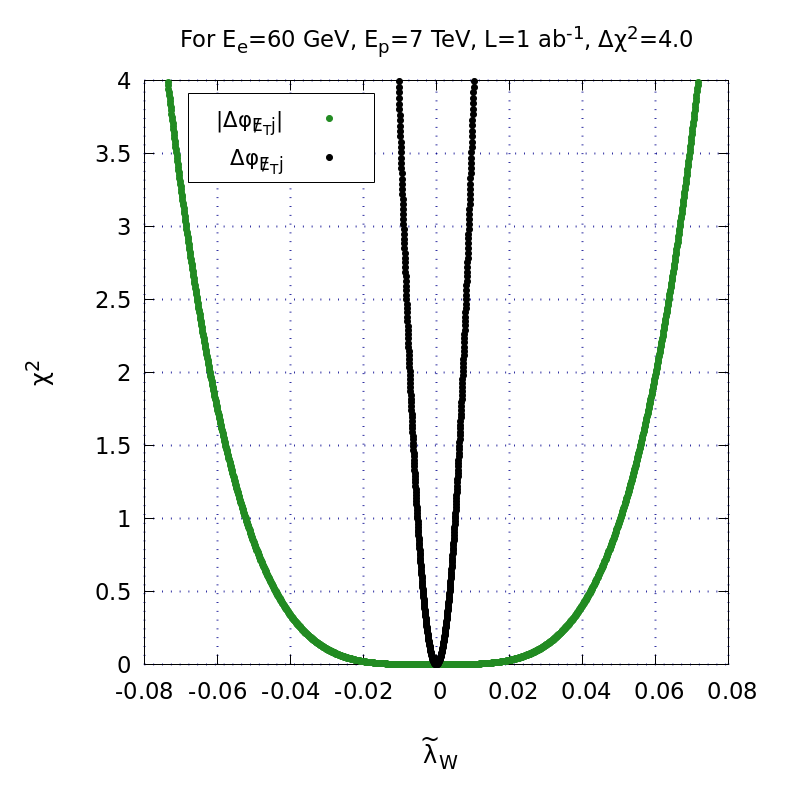}
    \label{chisq_ltw}
    }
	\caption{Constraints on non-standard $HZZ$ coupling parameters and ${\tilde \lambda}_W$ assuming $L=1$ ab$^{-1}$.}
	\label{fig:chisq}
\end{figure}

First, we take into account the scenario where only one of the three BSM parameters is taken to be non-zero at a time. The analysis considering more than one bin of an angular observable is expected to be more efficient than the one-bin analysis. We used a bin width of $9^{\circ}$ for all observables. This means 20 bins for the $\theta_e$ and $|\Delta \phi_{ej}|$ distributions, while 40 bins for $\Delta \phi_{ej}$.  Fig.~\ref{fig:chisq} presents $\chi^2$ plots assuming an integrated luminosity, $L=1$ ab$^{-1}$. The $\Delta \phi_{ej}$  distribution provides the strongest bound on $\lambda_{1Z}$, as shown in Fig.~\ref{chisq_l1}. 
In the case of $\lambda_{2Z}$, Fig.~\ref{chisq_l2} shows that the most stringent bounds are obtained from the $\theta_e$ distribution. 
For both the CP-even parameters $\lambda_{1Z}$ and $\lambda_{2Z}$, the constraints provided by $\Delta \phi_{ej}$ are slightly tighter than those obtained by $|\Delta \phi_{ej}|$. This is mainly due to the 
larger number of bins that go into the analysis using $\Delta \phi_{ej}$ distribution. Fig.~\ref{chisq_lt} demonstrates that the asymmetric distribution $\Delta \phi_{ej}$, which is sensitive to the CP-odd structure of the $HZZ$ coupling, puts the most stringent limits on $\widetilde{\lambda}Z$ compared to the other angular observables considered in this analysis. The expected effectiveness of $\Delta \phi _{\slashed{E}j}$ over $|\Delta \phi _{\slashed{E}j}|$ in constraining ${\tilde \lambda}_W$, analyzing the charged current process can also 
be seen in Fig.~\ref{chisq_ltw}.

\begin{table}[]
    \centering
    \begin{tabular}{c|c|c}
         Couplings &  $|\Delta \phi|$ & New observables \\
         \hline
        $\lambda_{1Z}$ &   [-0.07, 0.07]& [-0.07, 0.07] ($\Delta \phi_{ej}$)  \\
        $\lambda_{2Z}$ & [-0.02, 0.02] & [-0.01, 0.01] ($\theta_e$) \\
        $\widetilde{\lambda}_Z$ & [-0.22, 0.22] & [-0.07,0.07] ($\Delta\phi_{ej}$)\\
        \hline
    $\widetilde{\lambda}_W$ & [-0.07, 0.07] & [-0.01, 0.01] ($\Delta\phi_{\slashed{E}j}$)
    \end{tabular}
    \caption{Comparison of constraints obtained from $|\Delta \phi|$ (second column) with those from new observables $\Delta \phi$ and $\theta$ (third column).}
    \label{tab:ModDphiVsNew}
\end{table}

Using new observables, table~\ref{tab:ModDphiVsNew} shows that the constraints on $\lambda_{2Z}$ and $\widetilde{\lambda}_{Z}$ improve by 48\% and 67\%, respectively, compared to those obtained from the $|\Delta \phi_{ej}|$ distribution discussed in Ref.~\cite{Sharma:2022epc}. It is worth noting here that the HL-LHC projections \cite{Cepeda:2019klc,Boselli:2017pef} on $\lambda_{1Z}$, $\lambda_{2Z}$ and $\widetilde{\lambda}_Z$ are 1\%, 0.5\% and 12\% at 95\% C.L. with an integrated luminosity of 3 ab$^{-1}$.
In the table, we also compare constraints from $\Delta \phi _{\slashed{E}j}$ and $|\Delta \phi _{\slashed{E}j}|$ on CP-odd component $\widetilde{\lambda}_W$ of non-standard $HWW$ coupling. 
We find an improvement of about 85\% when using $\Delta \phi _{\slashed{E}j}$ over $|\Delta \phi _{\slashed{E}j}|$ in the analysis. 
All these numbers are expected to change with the inclusion of systematic uncertainties.

\begin{figure}

 \includegraphics[width=0.60\textwidth,height=0.40\textwidth]{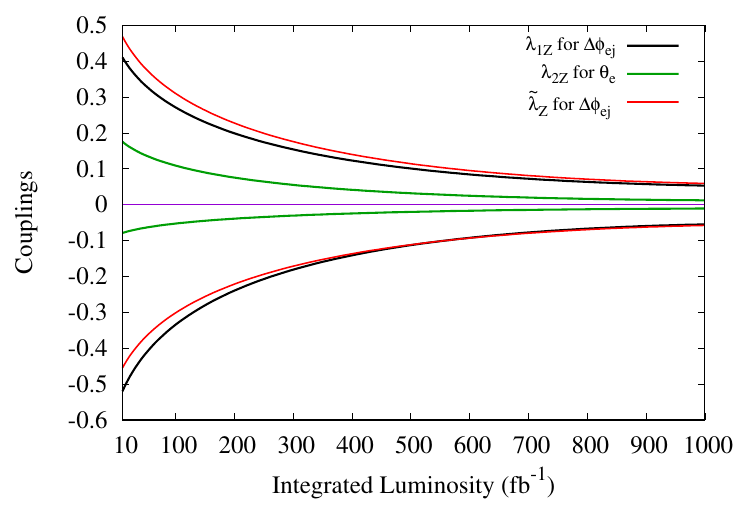}
	\caption{Variation of constraints on $HZZ$ parameters using the best observables obtained at 95\% C.L. as a function of integrated luminosity.}
	\label{fig:lumi}
\end{figure}

The constraints on $\lambda_{1Z}$, $\lambda_{2Z}$ and $\tilde{\lambda}_Z$ as a function of luminosity, varying from 10 fb$^{-1}$ to 1000 fb$^{-1}$, are presented in Fig.~\ref{fig:lumi} at 95 \% C.L. In this plot, constraints on each BSM parameter are taken from the most sensitive observable to that parameter. The variation of bounds on all couplings with luminosity follows a similar trend. 
 The constraints on $\lambda_{1Z}$, $\lambda_{2Z}$ and $\tilde{\lambda}_Z$ change by 74\%, 83\%, and 74\%, respectively, as the luminosity increases from  10 fb$^{-1}$ to $10^3$ fb$^{-1}$.

\begin{figure}[h!]	
\includegraphics[width=0.49\textwidth]{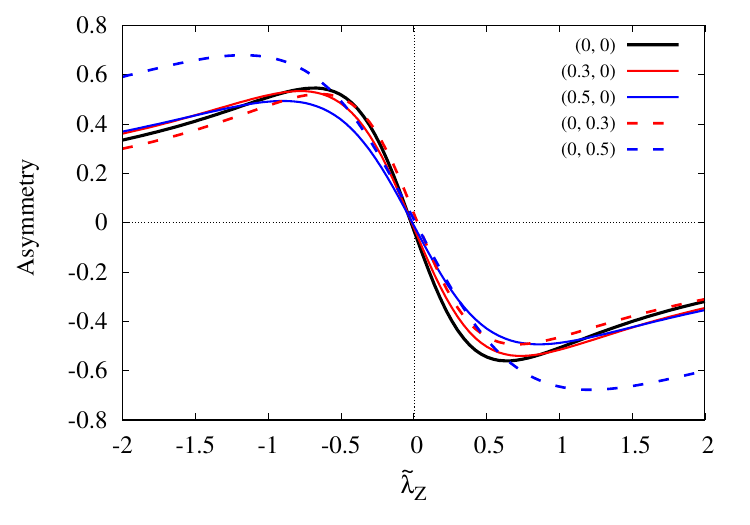}
\includegraphics[width=0.40\textwidth]{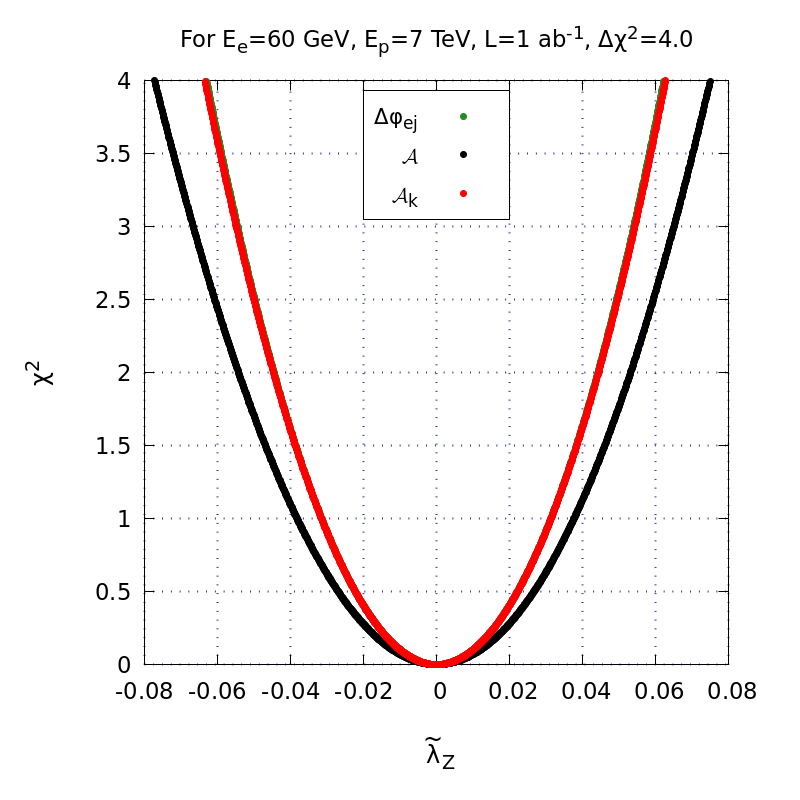}
	\caption{Asymmetry in $\Delta \phi_{ej}$ as a function of $\widetilde{\lambda}_Z$ for fixed values of CP-even couplings $(\lambda_{1Z},\lambda_{2Z}$) (left), and the corresponding $\chi^2$ plot based on asymmetry in $\Delta \phi_{ej}$ (right).}
	\label{AsymVsLt}
\end{figure}

The asymmetric nature of $\Delta \phi_{ej}$, as shown in Fig.~\ref{fig_obs:Dphi} for $\widetilde{\lambda}_Z$, allows us to construct an asymmetry observable that is less sensitive to systematic uncertainty. The asymmetry is defined as: 
\begin{equation}
    \mathcal{A} =
        \frac{\mathcal{O}_{\Delta\phi > 0} - \mathcal{O}_{\Delta\phi < 0}}{\mathcal{O}_{\Delta\phi > 0} + \mathcal{O}_{\Delta\phi < 0}}
        \label{obs_asym}         
\end{equation}
When using $\mathcal{A}$ in the $\chi^2$ analysis, the statistical uncertainty is given by
\begin{align*}
	\Delta \mathcal{O} = \sqrt{\frac{1-(\mathcal{A}^{\rm SM})^2}{\mathcal{O}^{\rm SM}.L}}. 
\end{align*}
Here, $L$ is the integrated luminosity.

Since the amplitude-squared is an even function of $\Delta \phi_{ej}$ for CP-even coupling and an odd function for CP-odd coupling, the asymmetry has a linear dependency on the CP-odd coupling in its numerator. However, its denominator can depend on all three couplings $\lambda_{1Z}$, $\lambda_{2Z}$ and $\widetilde{\lambda}_Z$. Consequently, for small values of CP-even couplings, the asymmetry varies linearly, as shown in Fig.~\ref{AsymVsLt}. This observable can be used to constrain the CP-odd coupling when the other parameters are tightly constrained by any theoretical or experimental study. Using this asymmetry in the $\chi^2$ formula given in Eq.~\ref{chiSq}, we first obtain constraints on $\widetilde{\lambda}_Z$ under the assumption that the other BSM parameters are zero. In Fig.~\ref{AsymVsLt}, we have plotted $\chi^2$ as function of ${\tilde \lambda}_Z$ using the asymmetry ${\cal A}$ defined in Eq.~\ref{obs_asym}, bin-wise asymmetry ${\cal A}_k$, and $\Delta \phi_{ej}$ as observables. The bin-wise asymmetry ${\cal A}_k$ is constructed by applying the definition of ${\cal A}$ for each pair of symmetrically located bins about $\Delta \phi_{ej}=0$. Thus, a total of 20 values of  ${\cal A}_k$ is 
used in the analysis. As expected, the constraints using ${\cal A}_k$ are tighter than those obtained using ${\cal A}$. However, there is no improvement in constraints when comparing them with the analysis based on $\Delta \phi_{ej}$. In the presence of systematic uncertainty, ${\cal A}_k$  is expected to be a better observable than $\Delta \phi_{ej}$. The above qualitative features remain true in the $\chi^2$ analysis of ${\tilde \lambda_W}$ parameter using the charged-current process.

\begin{figure}
	\includegraphics[width=0.40\textwidth]{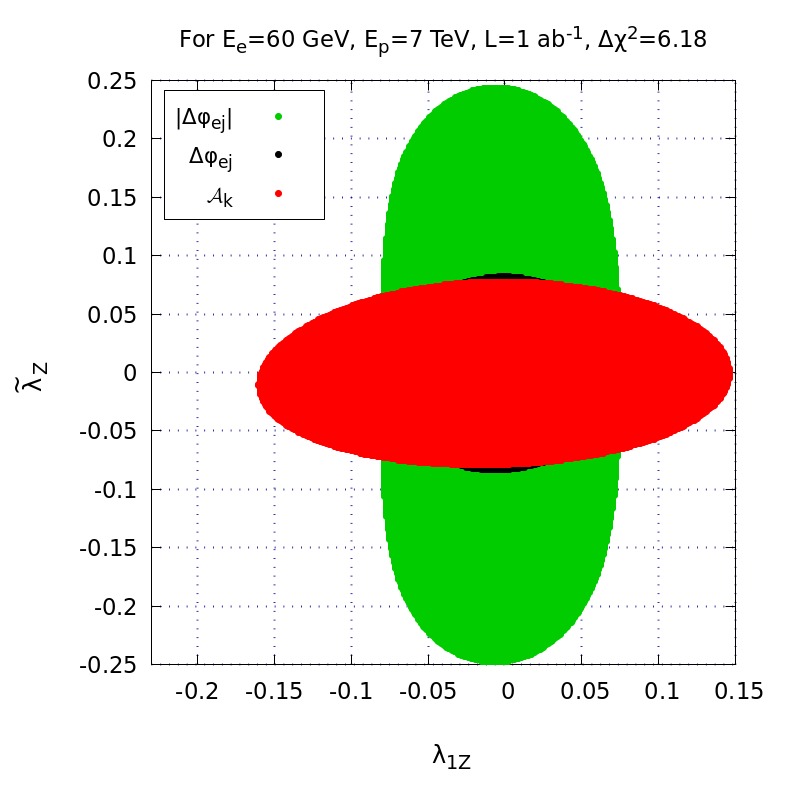}
 	\includegraphics[width=0.40\textwidth]{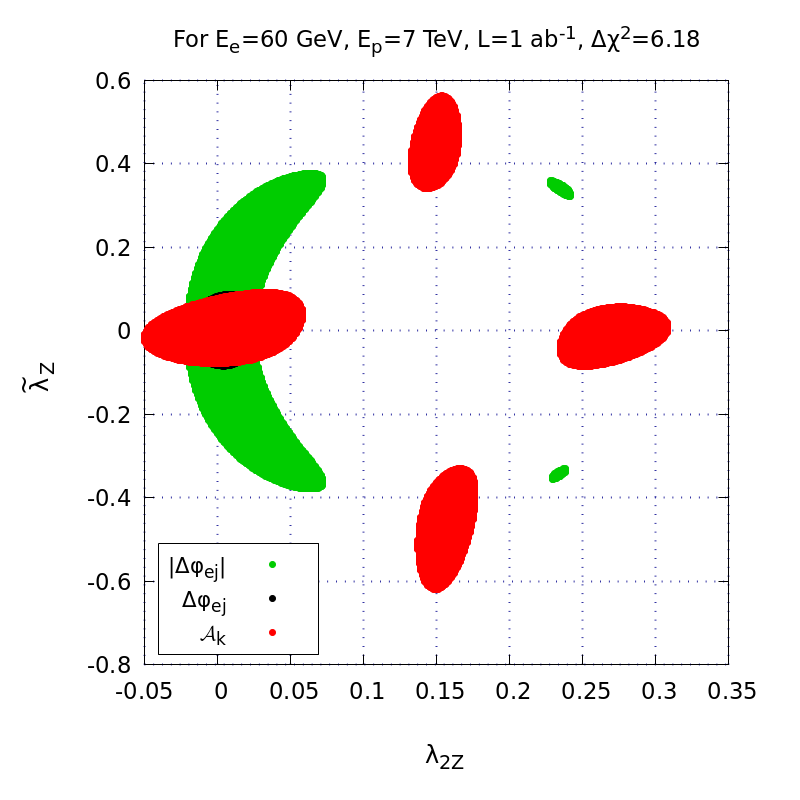}
	\caption{Comparison of constraints in the two parameter space based on $\Delta \phi_{ej}$, $|\Delta \phi_{ej}|$, and the asymmetry in $\Delta \phi_{ej}$.}
	\label{chi_2param}
\end{figure}

As discussed in the previous section, the $\Delta \phi_{ej}$ distribution is more efficient than the $|\Delta \phi_{ej}|$ distribution in capturing CP-odd coupling effects when CP-even couplings are present. Considering this fact, we present constraints on $\widetilde{\lambda}_Z$ in Fig.~\ref{chi_2param}, assuming the presence of one CP-even coupling at a time. From this figure, it is evident that the constraints on $\widetilde{\lambda}_Z$ are considerably stronger for $\Delta \phi_{ej}$ (black) than $|\Delta \phi_{ej}|$ (green). Around smaller values of CP-even couplings, the asymmetry (red) performs slightly better than $\Delta \phi_{ej}$. However, as discussed above, for larger values of $\lambda_{1Z}$ and $\lambda_{2Z}$ sensitivity of asymmetry is reduced compared to the angular distribution.

\section{Conclusions}
A general Lorentz-invariant $HZZ$ coupling, parameterized by the three parameters $\lambda_{1Z}$, $\lambda_{2Z}$, and $\tilde{\lambda}_Z$ is studied via  the \texttt{NC} process at an $ep$ collider. We analyze the sensitivity of various angular observables--$\theta_e$, $\alpha_{ej}$, $|\Delta \phi_{ej}|$, $\beta_{ejh}$, and $\Delta \phi_{ej}$--to each coupling by setting the remaining parameters to zero. 
It is found that $\Delta \phi_{ej}$ and $\theta_{e}$ are the most sensitive angular observables. We have compared the constraints on each parameter using these observables. The strongest bound on $\lambda_{1Z}$ comes from the $\Delta \phi_{ej}$ distribution, which is the same as obtained in our previous study \cite{Sharma:2022epc}. The observables $\theta_{e}$ and $\Delta \phi_{ej}$ provide strongest limits on the $\lambda_{2Z}$ and $\tilde{\lambda}_Z$ parameters, respectively. At an integrated luminosity of 1 ab$^{-1}$, these limits now fall within the ranges of [-0.02, 0.02] for $\lambda_{2Z}$  and [-0.22, 0.22] for $\tilde{\lambda}_Z$, representing improvements of 48\% and 67\%, respectively, compared to our previous study. The angular observable $\Delta \phi_{ej}$ shows an asymmetry around $\Delta \phi_{ej}$ = 0 in the presence of the CP-odd coupling $\widetilde{\lambda}_Z$, which arises due to the four-rank epsilon tensor structure. The constraints on $\widetilde{\lambda}_Z$ derived from this asymmetry are closely comparable to those obtained from the $\Delta \phi_{ej}$ distribution when the other parameters are set to zero. Furthermore, we present constraints on the CP-odd coupling in scenarios where one of the CP-even couplings is nonzero. In such cases, the asymmetry in the $\Delta \phi_{ej}$ distribution provides stronger constraints on $\widetilde{\lambda}_Z$ compared to the $\Delta \phi_{ej}$ observable when $\lambda_{1Z}$ is non-zero. However, $\Delta \phi_{ej}$ sets a tighter parameter space for $\widetilde{\lambda}_Z$ in the presence of $\lambda_{2Z}$. The constraints on $\widetilde{\lambda}_Z$ obtained from the new observable are stronger than the HL-LHC projections, whereas those on $\lambda_{1Z}$ and $\lambda_{2Z}$ are weaker for the achievable integrated luminosities at HL-LHC and LHeC. We have noted that like $\Delta \phi_{ej}$ in the 
neutral current process, $\Delta \phi_{\slashed{E}j}$ (ranging between $-\pi$ to $\pi$) can be constructed in the charged-current process without any ambiguity leading to an improved constraint on ${\tilde \lambda}_{W}$, the CP-odd parameter of $HWW$ coupling. 

\section{Acknowledgments}
{We would like to acknowledge fruitful discussions with Sudhansu S. Biswal and Soureek Mitra. PS would like to acknowledge financial support from IISER Mohali for this work.}

\newpage

\bibliographystyle{unsrt}
\bibliography{ref}

\end{document}